\PassOptionsToClass{table}{xcolor}
\documentclass[sigconf,natbib=false,noacm]{acmart}
\usepackage[utf8]{inputenc}
\usepackage[T1]{fontenc} % optional
\usepackage{amsmath,amsfonts}
\usepackage{bm}
\usepackage[scale=.9]{cascadia-code}
\usepackage{algorithmic}
\usepackage{graphicx}
\usepackage{textcomp}
\usepackage{xfp}
\usepackage{xcolor}
\usepackage{colortbl}
\usepackage[backend=biber,style=acmnumeric]{biblatex}
\AtBeginDocument{%
  }
\usepackage{url}
\definecolor{spaceorange}{HTML}{ED820E}

\usepackage{fontawesome}
\usepackage{tikz}
\usetikzlibrary{tikzmark,shapes.symbols}
\usepackage{makecell}
\usepackage{booktabs}
\usepackage{csquotes}
\PassOptionsToPackage{hidelinks}{hyperref}
\usepackage{hyperref}
\usepackage[capitalize,nameinlink,noabbrev]{cleveref}

\usepackage{microtype}
\usepackage[inline]{enumitem}
\newlist{inlist}{enumerate*}{1}
\setlist[inlist]{itemjoin={{, }},itemjoin*={{, and }},label=(\(\roman*\)),mode=boxed}
\usepackage{forest}
\useforestlibrary{edges}
\def\SWidth{3.25cm}%
\forestset{
   internal/.style={gray,font=\sffamily},
   marker/.style={rectangle split parts=1,minimum width=0cm},
   norm-ast-node/.style={
      Rect,
      outer sep=1pt,
      rounded corners=1pt,
      Soft,
      s sep=2pt,
      l sep+=2mm,
      rectangle split,
      rectangle split parts=2,
      font=\ttfamily,
   },
   l2r/.style={
      grow'=east,
      anchor = west,
      parent anchor = east,
      child anchor = west,
   },
   rot90/.style={
      rotate=90,grow=east,anchor=north,parent anchor=south,child anchor=north
   },
   norm-ast/.style={
      norm-ast-node,
      minimum width=\SWidth,
      l2r,
      line join=round,
   },
   o/.style={edge label={node[pos=.5,fill=white,inner sep=1.25pt,rounded corners=1pt,font=\scriptsize] {\strut #1}}},
   al-center/.style={
      before computing xy={s/.average={s}{siblings}}
   }
}
\def\NodeSurround{~~}%
\newsavebox\NodeTitleBox
\newcommand\Node[3][\SWidth]{\global\setbox\NodeTitleBox=\hbox{\NodeSurround#2\NodeSurround}\usebox\NodeTitleBox \nodepart{two} \edef\w{\ifdim\wd\NodeTitleBox<#1 #1\else\wd\NodeTitleBox\fi}\parbox{\w}{\smaller\sffamily#3}}%

\newlist{blocklist}{enumerate*}{1}
\setlist[blocklist]{itemjoin={{\space}},itemjoin*={{\space}},label=(\(\roman*\)),mode=boxed}
\newlist{orlist}{enumerate*}{1}
\setlist[orlist]{itemjoin={{, }},itemjoin*={{, or }},label=(\(\alph*\)),mode=boxed}
\usepackage{wrapfig}
\usetikzlibrary{arrows.meta,fit,backgrounds,calc}
\tikzset{
   Soft/.style={line join=round,line cap=round},
   All Soft/.style={every path/.append style={Soft}},
   Blob/.style={
      draw=black,
      circle,
      outer sep=2pt,
      minimum size=1.8em
   },
   Blobs/.style={
      every node/.append style={Blob}
   },
   Use/.style={Blob},
   RectRounding/.style={rounded corners=3pt},
   Rect/.style={
      draw=black,
      rectangle,
      RectRounding,
      outer sep=2pt,
      minimum size=1.8em
   },
   Def/.style={Rect},
   Rects/.style={
      every node/.append style={Rect,minimum width=#1}
   },
   Rects/.default={1.5em},
   Link/.style={
      draw,
      Soft,
      rounded corners=2pt,
      -Kite
   },
   Links/.style={
      every path/.append style={Link}
   },
   comm/.style={rectangle,draw,text width=7.5mm,align=center,minimum height=5mm,font=\small\ttfamily,fill=lightgray!22!white},
   d/.style={comm,rounded corners=2pt,hyperlink node={v:def}}, % variable def
   u/.style={comm,rounded corners=2.5mm,hyperlink node={v:use}}, % variable use % rounded rectangle breaks anchors
   v/.style={comm,signal, signal to=east and west, gray, text=black, inner sep=-4pt, text width=5mm,fill=lightgray!22!white,hyperlink node={v:value}}, % values
   F/.style={comm,draw=gray,rounded corners=2pt,fill=white,inner xsep=.5em,hyperlink node={v:fdef}},
   fc/.style={comm,double,rounded corners=2.5mm, outer sep=1.15pt,hyperlink node={v:call}},
   w-back/.style={fill=white,inner sep=1pt},
   T/.style={font=\footnotesize\ttfamily,text=gray},
   code/.style={font=\ttfamily},
   dfidn/.style={
      circle,darkgray,fill opacity=.925,xshift=-1mm,yshift=.15mm,fill=white,draw,minimum size=1.8em,scale=.8,inner sep=-1pt
   },
   klabel/.style={font=\scriptsize\sffamily, inner sep=1pt,text=black,above,execute at begin node={\strut}},
   olabel/.style={midway,klabel},
   pos at/.style={above #1=-.5mm,yshift=-.4\baselineskip},
   old/.style={opacity=.65},
   graph-frame-style/.style={very thick,rounded corners=1.5pt,draw,black},
   graph-cd/.style={gray,font=\scriptsize,line cap=round}, 
   graph-cd-in/.style={graph-cd,{Kite[scale=.7]}-},
   graph-cd-out/.style={graph-cd,-{Kite[scale=.7]}}
}
\tikzset{
    hyperlink node/.style={
        alias=sourcenode,
        append after command={
            let     \p1 = (sourcenode.north west),
                \p2=(sourcenode.south east),
                \n1={\x2-\x1},
                \n2={\y1-\y2} in
            node [inner sep=0pt, outer sep=0pt,anchor=north west,at=(\p1)] {\hyperlink{#1}{\XeTeXLinkBox{\phantom{\rule{\n1}{\n2}}}}}
        }
    }
}
\def\DefineEdgeType#1#2#3{%
   \expandafter\def\csname edge#1short\endcsname{#1}
   \expandafter\def\csname edge#1long\endcsname{#2}
   \expandafter\def\csname edge#1desc\endcsname{\def\s{\textit{source}\xspace}\def\t{\textit{target}\xspace}#3}
}
\usepackage{xspace}
\def\GetEdge#1#2{\csname edge#1#2\endcsname}

\DefineEdgeType{reads}{reads}{\s reads from \t}
\DefineEdgeType{def-by}{defined-by}{\s is defined by the \t}
\DefineEdgeType{calls}{calls}{\s calls the \t}
\DefineEdgeType{returns}{returns}{\s returns to \t}
\DefineEdgeType{def-on-call}{defines (on call)}{if the call occurs, \s~(argument) defines \t~(parameter)} % . This edge is bi-directional.
\DefineEdgeType{arg}{argument}{\t vertex is an argument of the \s}
\DefineEdgeType{side-effect}{side effect (on-call)}{if the call occurs, the \s effects on \t}
\DefineEdgeType{nse}{non-standard evaluation}{\s causes a non-standard evaluation of \t}
\RequirePackage[
   print,
   numinpar,
   fakeminted
]{lib/xlistings/xlistings}
\lstcolorlet{keywordA}{black}
\lstcolorlet{keywordB}{black}
\lstcolorlet{keywordC}{black}
\lstcolorlet{literals}{black}
\lstcolorlet{numbers}{darkgray}
\xlstDefineStyles{%
 {keywordA: \color{xlst@colors@lst@keywordA}\bfseries},%
 {keywordB: \color{xlst@colors@lst@keywordB}},%
 {keywordC: \color{xlst@colors@lst@keywordC}},%
 {keywordD: \itshape},%
 {numbers:  \color{xlst@colors@lst@numbers}\slshape},%
 {literals: \color{xlst@colors@lst@literals}\slshape},%
 {comments: \color{xlst@colors@lst@comments}},% \scshape
 {basic:   \ttfamily},%
 {command:  \color{xlst@colors@lst@command}}%
}
\LoadLanguage{sle-R}
\xlstsetmintedstyle{plain number}

\AtBeginDocument{\xlstInlinePreBreak{}\xlstPreBreak{}}
\RequirePackage{tabularx}
\RequirePackage[detect-all]{siunitx}
\usepackage{multicol}
\usepackage{xfp}
\usepackage{multirow}
\def\PrintResultOfThe#1#2#3#4{%
   \edef\result{\fpeval{round(#1,#2)}}%
   \edef\chk{\fpeval{\result=0?1:0}}%
   \ifnum\chk=1\relax
      \edef\numcmpres{\fpeval{1/(10^#2)}}%
      \qty{< \numcmpres}{#4}%
   \else
      #3\relax
   \fi
}
\newcommand*\ThePercent[3][2]{%
   \edef\fst{\fpeval{#2}}%
   \edef\snd{\fpeval{#3}}%
   \edef\result{\fpeval{\fst/\snd * 100}}%
   \PrintResultOfThe{\result}{#1}{\ifdim\result sp=100sp\relax\ifdim\fst sp=\snd sp\else\(\approx\)\fi\fi\qty{\fpeval{\result}}{\percent}}{\percent}%
}
\def\FAIR{\textsc{fair}\xspace}
\usepackage[subtle]{savetrees}

\begin{abstract}
   High-level languages such as~R or~Python are used frequently to analyze and visualize data in the form of scripts or notebooks. However, these artifacts suffer from reproducibility issues due to what we frame as implicit assumptions made by the authors. %  \mtt{haben wir nachher Evidenz dafür?}
   Such assumptions range from package versions and shapes of involved data tables, to manual and often undocumented setup steps.
   Within this work, we provide a unified, example-driven perspective on implicit assumptions in data analysis backed by an explorative proof-of-concept implementation. With this perspective, we propose the use of static analysis techniques to identify such assumptions and to make them explicit in the form of code constraints, focusing on the inclusion of data-analysis-specific issues. Such constraints can then be used to automatically transform these scripts into executable and reproducible artifacts, to check these assumptions at runtime, and to serve as documentation to support code reuse and comprehension.
\end{abstract}
 %  Programming Language

\begin{document}
\def\OurTitle{Towards Automatically Inferring Constraints to Identify Implicit Assumptions in Data Analysis}
\title{\OurTitle}
\author{Florian Sihler}
\email{florian.sihler@uni-ulm.de}
\orcid{0000-0001-7195-7801}
\affiliation{%
  \institution{Ulm University}
  \country{Germany}
}
\author{Lars Pfrenger}
\email{lars.pfrenger@uni-ulm.de}
\orcid{0009-0000-8166-7023}
\affiliation{%
  \institution{Ulm University}
  \country{Germany}
}
\author{Oliver Gerstl}
\email{oliver.gerstl@uni-ulm.de}
\orcid{0009-0007-5612-0780}
\affiliation{%
  \institution{Ulm University}
  \country{Germany}
}
\author{Matthias Tichy}
\email{matthias.tichy@uni-ulm.de}
\orcid{0000-0002-9067-3748}
\affiliation{%
  \institution{Ulm University}
  \country{Germany}
}

\setcopyright{none}
\acmConference[]{}{}{}
\acmBooktitle{}
\acmDOI{}
\acmISBN{}
\copyrightyear{}
\renewcommand\footnotetextcopyrightpermission[1]{}
\maketitle
\errorcontextlines=99999
\overfullrule=2cm
\section{Introduction}
There are many high-level data-analysis languages such as R, Python, MATLAB, and Julia which offer a sophisticated range of features for statistical tests and visualization~\cite{hadley2016r,embarak2018data,DBLP:conf/sc/PrabhuJRZHKJLGB11}. Additionally, these languages benefit from notebook environments such as Jupyter Notebook~\cite{DBLP:journals/cse/GrangerP21} or Quarto~\cite{Allaire_Quarto_2024}, which support an interactive exploration and manipulation of data. These technologies support domain experts in conducting their analyses without training in software engineering principles~\cite{DBLP:journals/rjour/Vidoni21}. 
In practice, however, analysis scripts~\cite{trisovic_largescale_2022,DBLP:conf/msr/SihlerPSTDD24}
and notebooks~\cite{DBLP:conf/msr/IslamAW24,DBLP:conf/msr/PimentelMBF19,10.1093/gigascience/giad113} suffer issues in reproducibility and structure~\cite{DBLP:conf/vl/0002HB20}. 
Various studies show that they lack executability~\cite{trisovic_largescale_2022,DBLP:conf/msr/PimentelMBF19,10.1093/gigascience/giad113,DBLP:conf/msr/IslamAW24} and hence reproducibility. These studies reveal that~\qtyrange{65}{77}{\percent} of the analyzed artifacts fail to execute due to missing dependencies, wrong execution order, or bad coding practices~\cite{trisovic_largescale_2022}. Moreover, research on code reuse in data science finds that~\qtyrange{70}{80}{\percent} of the analyzed code snippets are clones of other snippets with~\qty{50}{\percent} containing no unique code at all~\cite{10.7717/peerj.13933,DBLP:journals/programming/KallenW21}. Code reuse is even more prevalent across different projects. This indicates significant reproducibility issues of data science code.
Such artifacts are used frequently to help other researchers understand the applied methods and the steps conducted in the accompanied publication or to reuse parts within their own related analyses~\cite{DBLP:conf/vl/KoenzenES20,10.7717/peerj.13933}.
Reasons for these difficulties are manifold. Ranging from finding the used versions of packages and the language, to external dependencies like resource files. Additionally, they may rely on previous steps or expect the dataset to have a specific shape~\cite{trisovic_largescale_2022}.
While there is already some research on improving the reproducibility of data science scripts and notebooks, there is not yet a unified view on the various challenges that arise when trying to reproduce, replicate, or reuse these artifacts.
In this work, we address this by framing these difficulties as implicit assumptions, similar to classical software development~\cite{mamun2011review}. Additionally, we propose the systematic use of static analysis techniques to identify and document these implicit assumptions in the form of constraints, e.g., by using abstract interpretation to infer the values of variables~\cite{DBLP:conf/popl/CousotC77,10.1145/3689609.3689996}.
Such an analysis may happen either in the context of the original analysis environment~(cf.~\cite{donat2025r4r}) or retroactively when a researcher tries to reproduce, replicate, or reuse the artifact.
The results can then be used to add checks to the program that serve as an explicit documentation and as a dynamic validation for each run of the script.
Within the rest of this paper, we use the R~programming language as a primary example to illustrate specific cases. Furthermore, we use it as a basis for an explorative proof-of-concept implementation of our idea on top of \textit{flowR}, a static program analysis framework for~R~\cite{floopsla}. However, the proposed approach and most of the challenges are inherent to data analysis and hence language-agnostic. Therefore, we include a brief overview of the generalizability of the approach. Finally, we conclude with the implications of our work as well as a list of next steps in \cref{sec:conclusion,sec:future-plans}.

\section{Related Work}\label{sec:related-work}
To address issues of non-executable or non-reproducible research, several solutions have been proposed~(cf.~\cite{siddik2025slr,donat2025r4r}), including tools to gather executable code versions~\cite{DBLP:conf/chi/HeadHBDD19}, recreate the execution order of notebook cells~\cite{DBLP:conf/kbse/WangKLZ20}, provide linting recommendations~\cite{DBLP:journals/ese/PimentelMBF21}, infer the environment of Python code~\cite{DBLP:conf/icse/HortonP19,DBLP:conf/icse/Ye0DW022}, and restore the execution environment of notebooks~\cite{DBLP:conf/icse/WangLZ21}. Yet, neither do these tools address the potential interactions between assumptions, such as the version of an interpreter restricting the available libraries, nor do they provide a systematic approach to document these assumptions in the artifact.
One widely used technique for inferring code constraints is abstract interpretation introduced by \citeauthor{DBLP:conf/popl/CousotC77}~\cite{DBLP:conf/popl/CousotC77}.
It can be used to automatically infer preconditions on the code~\cite{DBLP:conf/vmcai/CousotCFL13} and statically check code contracts~\cite{DBLP:journals/computer/Meyer92,DBLP:conf/foveoos/FahndrichL10}, for example, to infer shapes of input data for data science scripts~\cite{DBLP:journals/corr/abs-2007-10688,urban:hal-04249957}. Additionally, subsequent research by \citeauthor{DBLP:conf/icse/SuboticMS22} presents a general static analysis framework for data science notebooks~\cite{DBLP:conf/icse/SuboticMS22}. Furthermore, \citeauthor{DBLP:conf/pldi/NegriniSU23} propose a static analyzer for data transformations in Jupyter notebooks~\cite{DBLP:conf/pldi/NegriniSU23} and \citeauthor{10.1145/3689609.3689996} introduce a high-level linter to identify code smells in data science scripts~\cite{10.1145/3689609.3689996}. However, none of these approaches focus on identifying and documenting issues of reproducibility, but instead they target the detection of code smells and visualization errors.

\def\sparagraph#1{\vspace{3pt}\par\smash{\raisebox{1pt}{\large\textbullet}}~\textit{#1.}~~\ignorespaces}
\section{Constraint Inference}\label{sec:overview}
This section explains our core idea of using constraint inference to identify implicit assumptions in data analysis scripts.
We propose to differentiate between two scenarios for constraint inference: A \textit{preemptive} analysis in the context of the original execution environment with the original data available, and \textit{retroactive} analysis when aspects of the original execution environment may be unknown or the original data may only be partially available.
In the former case, the researcher may want to ensure that the analysis script is self-contained and complies with the \FAIR principles~\cite{Wilkinson2016}. This case is usually easier as we can use the current computing environment to infer package versions and the shape of used datasets~\cite{donat2025r4r}.
For the latter case, we differentiate between three scenarios which benefit from the identification of implicit assumptions in research artifacts. A researcher may want to:
\begin{enumerate}[nosep]
  \item \textit{Reproduce} the results of the original study. %\todo{reproduce to back up study}
  \item \textit{Replicate} the results with a different dataset. % \todo{replicate with different data(set)}
  \item \textit{Reuse} parts within a different but related context~\cite{Wilkinson2016}. % \todo{re-use (maybe snippet) in a different context}
\end{enumerate}
We use this section to introduce the general idea with a running example~(\ref{sec:running-example}) and discuss the challenges alongside sketches of potential solutions of common assumptions like inferring package versions~(\ref{sec:package-versions}), script interdependencies~(\ref{sec:script-interdependencies}), and data shape expectations~(\ref{sec:data-shape-expectations}). % and the identification of manual setup steps~(\ref{sec:manual-setup}). % alongside their requirements for static and dynamic analysis.
\subsection{Running Example}\label{sec:running-example}
\def\lref#1{L\ref{#1}}% \mtt{vermutlich nicht real?}
Consider the following example R~script which reads a dataset from a \textsc{csv}~file~(\lref{c:read}),
calculates a mean score~(\lref{c:summ}) grouped by age~(\lref{c:group}), conducts a t-test~(\lref{c:ttest}), and plots the results~(\lref{c:plot1},\;\lref{c:plot2}). This follows the common data analysis workflow outlined by Wickham~\cite{hadley2016r}. Although the example is simple and artificial, it illustrates the common structure of data analysis scripts:
\begin{minted}[deletekeywords={t,c},escapeinside={|@}{@|}]{R}
|@\label{c:ggplot}@|library(ggplot2)
|@\label{c:read}@|da:c:ta <- read.csv("data.csv")
|@\label{c:age}@|b:c:y_age <- da:c:ta |>
|@\label{c:group}@|    dplyr::group_b:c:y(age) |>
|@\label{c:summ}@|    dplyr::summarise(m = mean(score))
|@\label{c:ttest}@|t.test(score ~ group, da:c:ta=grouped)
|@\label{c:plot1}@|ggplot(b:c:y_age, aes(x=age, y=m)) + 
|@\label{c:plot2}\label{c:last}@|    geom_c:c:ount()
\end{minted}
\def\expl#1#2{\textit{\color{gray}#1~\hfill#2}}
These \ref{c:last}~lines have the following ten implicit assumptions, which are common in real-world analyses~\cite{trisovic_largescale_2022,DBLP:conf/msr/IslamAW24,10.1093/gigascience/giad113}: % daniel mietchen, table 4 e.g.
\begin{enumerate}[label=A\arabic*,leftmargin=2em,ref=A\arabic*,nosep]
  \item \label{a:version}They require R~version 4.1 or newer\\*\expl{To support the native pipe operator: \texttt{|>}}{\lref{c:age},\;\ref{c:group}}
  \item \label{a:libraries}\texttt{ggplot2} and \texttt{dplyr} must be installed\\*
    \expl{Loaded as a library and accessed via \texttt{::}}{\lref{c:ggplot},\;\ref{c:group},\;\ref{c:summ}}
  \item \label{a:ggplot}\texttt{ggplot2} has to be at least version~2.0\\*
    \expl{This version introduces~\texttt{geom\_count}.\footnote{\label{fn:gg35r44}If we use R version 4.4 or later, we need \href{https://github.com/tidyverse/ggplot2/issues/6152}{ggplot2 version 3.5} or later.}}{\lref{c:plot2}}
  \item \label{a:datacsv}The file \textit{data.csv} must exist and be readable as a \textsc{csv}\\*
  \expl{Loaded by \texttt{read.csv}, relative to the working directory}{\lref{c:read}}
  \item \label{a:datacolumns}\textit{data} must have the columns \textit{age} and \textit{score}\\*
  \expl{They are used by \texttt{group\_by} and \texttt{summarise}}{\lref{c:group},\;\ref{c:summ}}
  \item \label{a:numeric}The \textit{score} column must be numeric\\*
  \expl{Or coercible to a number to work with \texttt{mean}}{\lref{c:summ}}
  \item \label{a:exist-in-env}The variable \textit{grouped} must exist in the environment\\*
  \expl{It is used in the t-test and not defined within the script}{\lref{c:ttest}}
  \item \label{a:datacolumns2}\textit{grouped} must contain a \textit{score} and \textit{group} column\\*
  \expl{Used in the formula of the t-test}{\lref{c:ttest}} 
  \item \label{a:levels}The \textit{group} column of \textit{grouped} must have two levels\\*
  \expl{The t-test requires a factor with exactly 2 levels}{\lref{c:ttest}}
  \item \label{a:assumptions}The values in \textit{grouped} conform to the t-test assumptions\\*
  \expl{Independence, normal distribution,~\ldots~\cite{delacre2017psychologists}}{\lref{c:ttest}}
\end{enumerate}
\noindent Some assumptions, like~\ref{a:assumptions}, are silent, meaning that they do not cause visible errors when violated but cause incorrect results, while others, like~\ref{a:libraries}, \textit{can} lead to runtime errors or warnings. However, they do not have to cause errors. For example, newer versions of \textit{ggplot2}, \textit{dplyr}, or~R may introduce changes in aesthetics, handling of edge-cases, value coercion, or deprecated functions, which can lead to different results.
Although it may be easy to identify these assumptions in the \ref{c:last}~example lines, this gets 
much harder for real-world scripts~\cite{DBLP:conf/msr/SihlerPSTDD24}. % \mtt{steht das dort?}.
In the context of our simple example, a preemptive analysis can get the exact version of used packages, knows whether \textit{data.csv} exists, the shape of the data, and more, just by reading the information from the environment and executing the script.
In contrast, a retroactive analysis is limited to inferring constraints about which package versions \textit{could} have been used and what the data \textit{might} have looked like.
Yet, this can still be important in detecting conflicts or reconstructing executable environments for the analysis.
\subsection[Inferring Package Versions]{Inferring Package Versions (\ref{a:version}--\ref{a:ggplot})}\label{sec:package-versions}
Incorrect versions are one of the main issues in reproducing data analyses~\cite{DBLP:journals/corr/abs-2303-04758}. There is a wide range of solutions that help to collect the used package versions and their dependencies preemptively.\footnote{See \href{https://cran.r-project.org/package=renv}{cran.r-project.org/package=renv}, \href{https://cran.r-project.org/package=groundhog}{cran.r-project.org/package=groundhog}.} Yet, work on reconstructing
such information retroactively is rare and either tries to obtain all package versions available at a given time~\cite{DBLP:journals/corr/abs-2303-04758} or analyze the API usage in notebooks~\cite{DBLP:conf/icse/WangLZ21}.  %  to infer version constraints
Currently, there exists no advanced and holistic approach to retroactively infer package versions using all of these information sources combined and document them in a machine readable format.
\sparagraph{Problem Statement}
Given an analysis artifact, combined with the available evaluation data and meta information, generate the requirements for the computing environment. These requirements include the required packages, their versions, and indirect dependencies such as system libraries.
If the original computing environment is available, this can be inferred preemptively~\cite{donat2025r4r}.
Hence, the main challenge is to infer these constraints retroactively.
\sparagraph{Proposed Solution}
To reconstruct the computing environment, we use a three-step approach: 
\begin{inlist}[itemjoin*={{, as well as }}]
  \item identify all dependencies required within the artifact
  \item uncover their potential origins and versions
  \item identify their system and interpreter constraints
\end{inlist}.
In~R, this includes loaded or attached libraries, those accessed via \enquote{\texttt{::}}~(\ref{a:libraries}), and those included indirectly~\cite{DBLP:journals/corr/abs-2303-04758} (for example, with \texttt{source} or as part of a \textsc{description} file).
For this we require more than call information. Common loading structures involve, e.g., mapping a vector of library names to the \texttt{library} function~\cite{DBLP:conf/msr/SihlerPSTDD24}. Hence, we need string and vector-aware domains to identify these patterns. Our proof-of-concept uses \textit{flowR}'s dataflow graph to identify these patterns and effectively infers all libraries requested by a given R~script, excluding only those with features not supported by \textit{flowR}~\cite{floopsla}. 
Next, the goal is to analyze the use of these packages within the artifact to infer possible package versions. Here we work on tracking \begin{inlist}
  \item called functions and their signatures
  \item used classes and global objects or constants
  \item reliance on loading effects such as the initialization of global variables
\end{inlist}.  
We can use this information to compare it with the available packages, their API, and dependencies to match up the expected behavior with the available versions~\cite{DBLP:conf/icse/WangLZ21}.
Considering only line~\ref{c:plot2} of the running example, this helps us to infer a minimum version constraint for \textit{ggplot2} based on the use of \texttt{geom\_count}~(\ref{a:version}). Incorporating \textit{ggplot2}'s dependency on the R~interpreter adds another constraint, requiring different versions based on the used version of R~(\ref{a:ggplot}).
To infer package versions, we are currently working on adding type inference, which allows us to infer function signatures. % These can then be used to match signatures with those extracted from specific package versions.
Using this strategy, we can build a constraint system describing all possible versions that may have been used in the original analysis.
This system can be enriched with additional information, such as a date range in which the study was conducted. % This narrows the possible versions of packages and the interpreter as is done by the \textit{groundhog} package~\cite{groundhog2025}.
If this process yields only a single combination of possible versions and their dependencies, we can use these to generate a corresponding virtual environment.
If multiple sets are possible, we can use the constraints to generate a set of possible environments to reproduce the analysis. This can then be combined with dynamic analysis, comparing the execution results, to identify the most likely environment fully automatically.
\sparagraph{Generalization} % Inferring Package Versions
Assumptions like \ref{a:version}--\ref{a:ggplot} are also common problems in other programming languages such as Julia or Python.
However, inferring the package versions is not only useful in hindsight.
These techniques can also be applied in the context of the original computing environment or in scenarios where a machine-readable format of the computing environment is available.
In such cases, the constraints can be used to check the compatibility of the stated requirements with the actual environment and uncover conflicts in the used API.
\subsection[Script Interdependencies]{Script Interdependencies (\ref{a:datacsv} and \ref{a:exist-in-env})}\label{sec:script-interdependencies}
Usually, research artifacts rely on multiple scripts or notebooks to conduct all of their analyses. However, similar to how individual notebook cells may depend on each other~\cite{DBLP:conf/chi/HeadHBDD19}, these scripts tend to have implicit dependencies on each other. One script may produce the data that another script consumes~(e.g.,~\ref{a:datacsv}), or clean the existing data as a side-effect. In a language like~R, scripts can even pass on variables if they are sourced in the same session. This can lead to cases like the use of \textit{grouped} in the running example~(\ref{sec:running-example}) which is not defined in the script itself~(e.g.~\ref{a:exist-in-env}). %  but may originate from another script
While there exist some rudimentary solutions which identify global variables,\footnote{\href{https://cran.r-project.org/package=globals}{cran.r-project.org/package=globals}} they are restricted to single sources and do not incorporate any potential script interaction, files, or package dependencies.
Similarly, while common notebook formats like Jupyter Notebooks or RMarkdown store the last output of a cell, they offer no way to automatically record artifact-wide relations or necessary re-executions of cells.
\sparagraph{Problem Statement}
Given an analysis artifact and the available execution environment~(\ref{sec:package-versions}), identify implicit dependencies between the scripts and their code.
Such dependencies may be enriched with the information about which scripts need to run within the same environment. % indicate simple order constraints for the execution, that 
Then, using these constraints, identify all potential execution orders, and their interactions. This is especially important in the case of ambiguity, in which a number of either scripts or cells could be executed first, but their effect on the global state is different. 
\sparagraph{Proposed Solution}
To identify dependencies between these scripts in the retroactive case, we propose a two-step approach. First, we build the program dependency graph of each script independently but follow explicit script inclusions (e.g.,\href{https://www.rdocumentation.org/packages/base/versions/3.6.2/topics/source}{\texttt{source}} calls).
With this, we automatically gain information about explicit dependencies, potentially undefined values and functions, globally defined identifiers, and files which are either written or read by the scripts.
In the second step, we can use this information to create links between scripts. % explicitly sourced
This also connects previously undefined variables and function calls with definitions of their identifiers, the loading of libraries to imported  functions, library loads to their installation~(\ref{sec:package-versions}), and file reads with their creation.
The resulting graph indicates: 
\begin{enumerate}[nosep,leftmargin=*]
  \item Scripts without any implicit dependency on or to other scripts. Such scripts may be executed completely independently.
  \item Scripts that may depend on other scripts. Those scripts should be executed before their dependents. The possible execution orders are the topological graph orderings.
  \item Scripts that still have global variables or functions which are undefined, or that read or write files that do not exist. % This may hint at required manual setup steps~(\ref{sec:manual-setup}) or missing data files.
\end{enumerate}
Granularity wise, this can also be applied to code cells in notebooks or even individual code-blocks such as function definitions.
We have started to implement this approach in \textit{flowR}~\cite{floopsla} and evaluated it on \num{8160}~real-world R~projects scraped from Zenodo, Havard Dataverse, JSS, and Figshare, with a total of \num{33336}~files. With this, we find that~\qty[round-precision=1]{\fpeval{1194/8160 * 100}}{\percent} of projects contain implicit interdependencies. % By detecting and resolving these dependencies, we can now ensure that scripts are executed in the correct order.
In the preemptive case, these dependencies can be solved by recording the execution order, including re-executions.
Existing tools such as \texttt{\href{https://rdrr.io/cran/CodeDepends/man/historyAsScript.html}{history\-As\-Script}},\footnote{\href{https://cran.r-project.org/package=CodeDepends}{cran.r-project.org/package=CodeDepends}} fail to incorporate project-wide dependencies or multiple sessions.
Hence, we propose to extend these tools to incorporate a project-wide view of the dependencies.
\sparagraph{Generalization} % Script Interdependencies
The concept of interdependencies of research scripts (assumptions like~\ref{a:datacsv}
or~\ref{a:exist-in-env}) or notebooks exists for every granularity and essentially every language in the form of a reliance on external state like the connection to a database. % \mtt{ist das zum Problem, der Lösung oder beides?}
Detecting such potential interdependencies is also helpful to ensure that an analysis does not rely on the execution order of the scripts or any external state. 
\subsection{Data Shape Expectations (\ref{a:datacolumns}, \ref{a:numeric}, \ref{a:datacolumns2}--\ref{a:assumptions})}\label{sec:data-shape-expectations}
In the real world, no data analysis artifact contains any kind of \enquote{unit tests} and quality assurance is rare~\cite{DBLP:conf/msr/SihlerPSTDD24,DBLP:journals/rjour/Vidoni21}. Similarly, these scripts almost never validate that loaded data conforms to the expected shape.
They simply access the columns they \enquote{know} or expect to exist~(\ref{a:datacolumns2}) and assume that the data is in the correct format~(\ref{a:numeric}).
Some of these assumptions are checked implicitly by the language or function that is called. %For example, by throwing an error if a requested column does not exist.
However, there is no way to detect that, for example, an assignment \textit{intends} to overwrite a column, that two measurements are assumed to be independent, or that the values should follow a normal distribution.
Some explorative work already tries to leverage abstract interpretation to infer series types to detect high-level errors like the wrong plot type~\cite{10.1145/3689609.3689996}, and to infer types for visualization~\cite{DBLP:conf/kbse/XiaLBC024}. More recent work also focuses on the dynamic postulation and verification of statistical assumptions, indicating the relevance of this work~\cite{10.1145/3729391}. Yet, there is no sophisticated approach to infer such implicit assumptions statically from the artifact itself and express them explicitly.
However, such guards are integral to support the reuse of analysis artifacts in different contexts (e.g., newer versions of the dataset).
\sparagraph{Problem Statement}
Take an analysis artifact including all available files and infer the constraints that loaded datasets have to fulfill. % \mtt{zu redundant?}
Then, semi-automatically insert checks into the artifact to dynamically ensure that the loaded data conforms to the expected shape.
\sparagraph{Proposed Solution}
We propose using a new set of abstract domains~\cite{cousot2021principles} to infer and track the shape of the data as it is loaded and manipulated by the scripts.
This includes the identification of the data shapes (e.g., the number and names of the columns and their types), the constraints on the data (e.g., that values must be unique or follow a normal distribution), and the functions used with the data.
This way, we can infer the expectations of functions, such as the requirements for a t-test~(\ref{a:assumptions}) or expected data shapes~(\ref{a:datacolumns}), and add guards to the artifact to dynamically ensure that the preconditions and expectations are met.
Moreover, we propose a semi-automatic approach to add additional constraints that can not be inferred directly from the data, but that a researcher may add in a preemptive scenario (e.g., that the result of a function must lie within a given interval).
So far, we added a data frame shape inference using abstract interpretation to \textit{flowR}. We define an abstract domain to track column names, the number of columns, and rows of data frames. For this, we map concrete data frame functions to a set of abstract operations. A first evaluation of the implementation shows promising results for tracking the shape data in analysis scripts.
Yet, this requires a careful study of common functions to identify useful constraints that can be inferred from the data and the analysis artifact.
Additionally, not every constraint can be directly inferred from the data (e.g., the independence of two measurements). % Hence, some have to be added semi-automatically by a researcher knowledgable of the study.
\sparagraph{Generalization}
Expectations on the shape of data are frequent and neither limited to data analysis nor to~R~\cite{DBLP:conf/pldi/NegriniSU23,10.1145/3729391} sharing similarities to, e.g., taint tracking and type inference.
Inferring the shape of data by tracking a known state and the operations on the data, can be applied to most scenarios which involve external data that may change over time.
Similarly, the inferred constraints can also help with comprehension, e.g., by uncovering unexpected requirements.

\section{Conclusion}\label{sec:conclusion}
Within this work, we introduce the idea of \textit{implicit assumptions} as a novel, unified perspective on the difficulties in reproducing, replicating, and reusing data analysis artifacts. 
Based on this perspective, we envision a set of techniques to address these difficulties with the help of static code analysis, and to enrich these artifacts with the necessary information to dynamically ensure their reproducibility and reusability. We combine this with an explorative implementation on top of \textit{flowR} which underlines the feasibility of our approach.
Yet, these categories are by no means exhaustive. For example, research artifacts may rely on undocumented, manual setup steps or interactivity from the user,
which hinder automated execution, and reduce the reproducibility of the
results.
Furthermore, such techniques are no universal solution as inferring these implicit assumptions is inherently limited by Rice's theorem~\cite{rice1953classes}.
Hence, we require thorough analyses of reproducible and irreproducible artifacts to identify common patterns to use as heuristics for these techniques. 
Additionally, we need to consider a human-in-the-loop approach if the detection of an implicit assumption works, but can not be automatically resolved.
However, even in these cases, we believe that the mere detection of implicit assumptions is still a valuable contribution, offering an interesting and valuable research direction.
\section{Future Plans}\label{sec:future-plans}
So far, we have already started working on a proof-of-concept implementation of our idea on top of \textit{flowR}~\cite{floopsla} and intend to complete this implementation by adding best-effort support for the detection of all implicit assumptions mentioned in \cref{sec:overview}.
To allow for additional constraints by users, we intend on adding a combination of a domain-specific language and the ability to detect violations of common assumptions as with \href{https://rdrr.io/r/base/stopifnot.html}{\texttt{stopifnot}} in~R. 
Subsequently, we plan to evaluate the usefulness of our approach in three ways:
\begin{itemize}[leftmargin=*,nosep]
   \item Take non-executable artifacts from public repositories and published studies~(e.g., those found by~\citeauthor{trisovic_largescale_2022}~\cite{trisovic_largescale_2022}) and check how many of them can be made executable with our approach, comparing this to the related work~(cf.~\cite{trisovic_largescale_2022,DBLP:conf/msr/IslamAW24,donat2025r4r}).
   \item Take executable artifacts and use fuzzing on their datasets to produce problematic and unproblematic variations and to get an understanding of the effectiveness of the inferred constraints to prevent problems in reusing the artifact with different data.
   \item Conduct a user study with R users to evaluate the usefulness of the inferred constraints as documentation for understanding and reusing the artifact by measuring their performance in understanding and reusing the artifact with and without the constraints.
\end{itemize}

\begin{acks}
   This work was supported by the \grantsponsor{dfg}{German Research Foundation (DFG)}{https://gepris.dfg.de/gepris/OCTOPUS}: \textit{\grantnum{dfg}{504226141}}. We extend our thanks to all \textit{flowR} contributors, especially Julian Schubert and Johanna Scheck. Additionally, we thank Jan Vitek and all of the anonymous reviewers for their invaluable feedback.
\end{acks}

\clearpage
\def\bibfont{\fontsize{7.5}{8}\selectfont}
\printbibliography
\end{document}